\documentclass[aps,prb,twocolumn]{revtex4}
\usepackage{graphicx}
\usepackage{epsfig}
\usepackage{amsmath}
\usepackage{graphics}
\usepackage{epsfig}
\usepackage{tikz}
\usetikzlibrary{decorations.pathmorphing}

\newcommand{\bra}[1]{\ensuremath{\langle{#1}|\,}}
\newcommand{\ket}[1]{\ensuremath{\,|{#1}\rangle}}

\begin{document}

\title{Non-renewal statistics for electron transport in  a molecular junction with  electron-vibration interaction}
\author{Daniel S. Kosov}
\address{College of Science and Engineering, James Cook University, Townsville, QLD, 4811, Australia 
}


\begin{abstract}
Quantum transport of electrons through a molecule is a series of individual electron tunneling events separated by stochastic waiting time intervals.
 We study the emergence of  temporal correlations between successive waiting times for the electron transport
in a vibrating molecular junction.  Using master equation approach, we compute 
 joint probability distribution for waiting times of two successive tunneling events. We show that  
  the probability distribution is completely reset after each tunneling event if molecular vibrations are  thermally equilibrated. If we treat vibrational dynamics exactly without imposing the equilibration constraint, the statistics of electron tunneling events become non-renewal. Non-renewal statistics between two waiting times $\tau_1$ and $\tau_2$ means that the density matrix of the molecule is not fully renewed after  time $\tau_1$ and the probability of observing waiting time  $\tau_2$  for the second electron transfer  depends on the previous electron waiting time $\tau_1$. The strong electron-vibration coupling is required for the emergence of the non-renewal statistics. We show that in  Franck-Condon blockade regime the extremely rare tunneling events become positively correlated. 
\end{abstract}

\maketitle
\section{Introduction}

On the microscopic quantum mechanical level, electron current  consists of a sequence of single electron tunneling events separated by random waiting time intervals. \cite{nazarov-book} Statistics of these waiting time intervals reveals the wealth of interesting information about details of quantum transport.  \cite{flindt13,sothmann14,flindt14,flindt17,wtd-transient,PhysRevB.92.125435,harbola15,rudge16a, rudge16b,kosov17-wtd} The statistical properties of the waiting times are usually studied using waiting time distribution (WTD), which is a  conditional  probability distribution 
 that  we observe  the electron transfer in the detector electrode (drain or source)  at time $t +\tau$ given that an electron was detected in the same electrode   at  time $t$.\cite{brandes08}
 WTD is a complementary  to very popular full counting statistics in quantum transport and it has recently gained  a significant popularity in the study of nanoscale and mesoscale systems.\cite{buttiker12,flindt13,sothmann14,flindt14,flindt15,wtd-transient,PhysRevB.92.125435,harbola15, rudge16a, rudge16b,kosov17-wtd}

The question which we discuss in this paper is the following. When an electron transfers through a molecular junction, is there exists a correlation between waiting times for successive electron tunneling or they are statistically independent? Is it possible to have, for example, a situation, when the second tunneling electron senses the waiting time of the previous electron and changes its own waiting time accordingly? Intuitively, we expect that such kind of statistical temporal correlations can emerge in molecular junctions with strong electron-vibrational coupling -- the excitation of a  particular vibrational state depends on the waiting time of the electron, and the waiting time for the second electron feels the previous one through this vibration. Strong coupling between electronic and nuclear dynamics distinguishes molecular junctions from other nanoscale quantum transport systems. The interplay between nuclear and electronic dynamics has already led to the discovery of distinctly molecular junction phenomena as Franck-Condon blockade, \cite{fcblockade05,PhysRevB.73.155306,doi:10.1021/acs.nanolett.5b03434} negative differential resistance, \cite{PhysRevB.83.115414,kuznetsov-ndr,galperin05,zazunov06,solvent12}  nonequilibrium  chemical reactions,\cite{dzhioev11,PhysRevB.86.195419,catalysis12} cooling of 
nuclear motion by electric current.\cite{galperin09,Ioffe:2008aa,PhysRevB.83.115414}

The basis for our work is the extension of ideas of WTD beyond studying statistics of single waiting time to the domain of multiple waiting times joint probability 
distribution.\cite{PhysRevB.95.045306} Renewal theory assumes that successive waiting times between transport events are statistically independent equally distributed random variables. 
 In this case the joint probability density of two successive waiting times $w_2(\tau_2, \tau_1)$ can be factorised into a product of two single-time distributions $w_1(\tau_2) w_1(\tau_1)$, that means that the distribution is totally "renewed" after  waiting time $\tau_1$. The non-renewal statistics means the existence of temporal correlations between the subsequent tunneling events $w_2(\tau_2, \tau_1) \ne w_1(\tau_2) w_1(\tau_1)$. The study of non-renewal statistics is a very interesting topic on its own. Although the questions of temporal correlations between electron tunneling events are only started to appear in quantum transport,\cite{PhysRevB.95.045306,flindt15} the non-renewal statistics has a long history in chemical physics where it was used to describe single-molecule processes in
spectroscopy\cite{nonrenewal-cao,PhysRevA.83.063841,nonrenewal-budini, doi:10.1021/jp061471w} and kinetics.\cite{nonrenewal-enzyme,cao08}

The paper is organised as follows. Section II overviews the derivation of the master equation for electron transport through a vibrating molecular junction. In Section III, we introduce quantum jump operators and derive the expression for the joint probability density of two successive waiting times, $w_2(\tau_1, \tau_2)$. Section IV describes the results of numerical  and analytical calculations. Section V summarises the main results of the paper.

We use natural  units for quantum transport  throughout the paper: $\hbar=k_{B}=e=1$.

\section{Master equation in the polaronic regime}

The single-molecular junction is a  molecule connected to macroscopic source (S) and  drain (D) electrodes. 
The corresponding Hamiltonian is
\begin{equation}
H= H_{\text{molecule}} + H_{\text{electrodes}} + H_T.
\end{equation}
The molecule is modelled by Anderson-Holstein model  -- a single electronic level  interacting with a localised vibration.
The molecular Hamiltonian is:
\begin{equation}
H_{\text{molecule}}= \epsilon_0 a^\dag a  + \lambda \omega (b^\dag + b) a^\dag a +  \omega b^\dag b,
\end{equation}
where $\epsilon_0$ is molecular orbital energy,  $\omega$ is molecular vibration energy, and $\lambda$ is the strength of the electron-vibration coupling. $a^\dagger (a) $  creates (annihilates) an electron on molecular orbital, and $b^+ (b)$ is bosonic creation (annihilation) operator for the molecular vibration. The electronic spin does not play any  role here and will not be included explicitly into the equations.
Electrodes have noninteracting electrons:
\begin{eqnarray}
H_{\text{electrodes}}=  \sum_{k,\alpha=S,D} \epsilon_{k\alpha} a^\dag_{k \alpha} a_{k \alpha},
\end{eqnarray}
where $a^\dagger_{k\alpha}$   creates  an electron in the single-particle state $k$ of the source(drain) electrode  $\alpha=S(D)$ and $a_{k\alpha}$ is the corresponding electron  annihilation operator.   The molecule-electrode coupling is described by  tunneling interaction
\begin{eqnarray}
H_T=  \sum_{k,  \alpha=S,D } t_\alpha (  a^\dag_{k \alpha} a  + a^\dag a_{k \alpha}),
\end{eqnarray}
where $t_\alpha$ is the tunneling matrix element.

Using Born-Markov approximation and Lang-Firsov transormation\cite{lang_firsov1963} we obtain the master equation \cite{PhysRevB.69.245302}:
\begin{widetext}
\begin{eqnarray}
\dot P_{0q}(t) &=& \sum_{\alpha q'} \Gamma^\alpha_{0q,1q'} P_{1q'} (t)  -  \Gamma^\alpha_{1q', 0q} P_{0q}(t),
\label{me1}
\\
\dot P_{1q}(t)&=& \sum_{\alpha q'} \Gamma^\alpha_{1q,0q'} P_{0q'}(t)  -  \Gamma^\alpha_{0q',1q} P_{1q}(t),
\label{me2}
\end{eqnarray}
\end{widetext}
where  $P_{nq}(t)$ is the probability that the molecule is occupied by $n$ electrons and  $q$ vibrational quanta at time $t$. 
The transition rates rates are:\cite{PhysRevB.69.245302}
\begin{equation}
\Gamma^\alpha_{0q',1q} =  \gamma^\alpha |X_{q'q}|^2 \left(1-f_\alpha[\epsilon-\omega (q'-q)] \right).
\end{equation}
--  transition from state occupied by one electron and $q$ vibrations to the electronically unoccupied state with $q'$ vibrations  by the electron transfer from the molecule to $\alpha=S,D$  electrode
and
\begin{equation} 
\Gamma^\alpha_{1q',0q} =    \gamma^\alpha |X_{q'q}|^2  f_\alpha[\epsilon+\omega (q'-q)]
\end{equation}
--  transition when electron is transferred from $\alpha$  electrode into the originally empty molecules simultaneously changing the vibrational state from $q$ to $q'$.
The rates depend on the occupation of electrodes  given by Fermi-Dirac numbers
\begin{equation}
f_\alpha(E) = \frac{1}{1+e^{(E-\mu_\alpha)/T}},
\end{equation}
where $T$ is the temperature and $\mu_\alpha$ is the chemical potential of the electrode $\alpha$. The rates also depend on the 
Franck-Condon factor
\begin{equation}
X_{qq'}= \bra{q} e^{-\lambda (b^\dag  -b)} \ket{q'},
\end{equation}
and the electronic level broadening 
\begin{equation}
\gamma^\alpha = 2 \pi t_\alpha ^2 \rho_\alpha
\end{equation}
where $\rho_\alpha$  is density of states in the electrode $\alpha$ taken at molecular orbital energy $\epsilon$.

\section{Quantum jumps operators for electron tunneling and  waiting time distributions}

We introduce probability vector ordered in such a way  that the electronic  probabilities enter in pairs for each vibrational  states
\begin{eqnarray}
\label{P}
\mathbf P(t) =
\begin{bmatrix}
P_{00} (t)\\
P_{10}(t)\\
{P_{01}(t)}\\
{P_{11}(t)}\\
\vdots
\\
{P_{0N}(t)}\\
{P_{1N}(t)}\\
\end{bmatrix} ,
\end{eqnarray}
where $N$ is the total number of vibrational states  included into the calculations.
We also define the identity  vector of length $2N$:
\begin{eqnarray}
\mathbf I =
\begin{bmatrix}
1\\
1\\
1\\
1\\
\vdots
\\
1\\
1\\
\end{bmatrix} .
\end{eqnarray}
The normalisation of the probability is given by the scalar product between $\mathbf I $ and $\mathbf P $ vectors $ ( \mathbf I,  \mathbf P(t))$ 
\begin{equation}
 ( \mathbf I,  \mathbf P(t))= \sum_{q=0}^N P_{0q}(t) + P_{1q}(t)  =1.
\end{equation}

Using this probability vector we write the master equation (\ref{me1},\ref{me2})  in the matrix form
\begin{equation}
\dot {\mathbf P}(t) = {\cal L} \mathbf P(t),
\label{master-eq}
\end{equation}
where $\cal L$ is the Liouvillian operator. 
The quantum jump operator is $2N\times 2N$   matrix which is defined through the actions on the probability vector:\cite{kosov17-wtd}
\begin{equation}
({ J} \mathbf P(t) )_{mq} =  \delta_{m0} \sum_{q'} \Gamma^{D}_{0q,1q'} P_{1q'} (t). 
\label{jump}
\end{equation}
It  describes the  tunneling of electron from the molecule to the drain electrode.

We assume that 
 the system has reached the nonequilibrium steady state. Therefore it is described by the steady state density matrix, which is the null vector of the full Liouvillian  
\begin{equation}
  {\cal L} \; \mathbf P =0.
  \label{ss}
\end{equation}
Let us begin to monitor time delays between sequential quantum tunnelings in the nonequilibrium steady state. 
WTD for two waiting times, $w_2(\tau_2,\tau_1) $, is defined as joint probability distribution that the first electron waits time  $\tau_1$ and  the next electron waits time $\tau_2$ for the tunneling to the drain electrode
\begin{eqnarray}
w_2(\tau_2,\tau_1) =
 ( \mathbf I, \; J \;  e^{ ({\cal L} -J ) \tau_2}   \;J  \; e^{ ({\cal L} - J ) \tau_1} \; J  \; \mathbf P ).
\label{wtd2-1}
\end{eqnarray}
The definition becomes physically obvious if one reads it from right to left:  The system is in the steady state described by the probability vector $\mathbf P$, then it undergoes quantum jump $J$, then idle without the quantum jump for time $\tau_1$, then again undergoes quantum jump $J$, idle  for time $\tau_2$ and then experiences the quantum jump $J$.
WTD for single waiting time  between two consecutive tunneling events is
\begin{eqnarray}
w_1(\tau) =
 ( \mathbf I, \; J  \; e^{ ({\cal L} - J ) \tau} \; J  \; \mathbf P ).
\label{wtd1-1}
\end{eqnarray}
and again this definition is quite self-explanatory.
Let us normalise these distributions
\begin{eqnarray}
&&\int_0^{\infty} d \tau_1 \int_0^{\infty} d \tau_2 \; w_2(\tau_2, \tau_1) 
\nonumber
\\
&&
=( \mathbf I, \; J \; ({\cal L} -J )^{-1}   \; J \; ({\cal L} -J )^{-1} \; J  \; \mathbf P )
\nonumber
\\
&&
=( \mathbf I, \;  (J -{\cal L} + {\cal L})  \;  ({\cal L} -J )^{-1}   \; J \; ({\cal L} -J )^{-1} \; J  \; \mathbf P )
\nonumber
\\
&&
=-( \mathbf I, \; J \; ({\cal L} -J )^{-1} \; J  \; \mathbf P )
\nonumber
\\
&&
=-( \mathbf I, \;  (J -{\cal L} + {\cal L})  \; ({\cal L} -J )^{-1} \; J  \; \mathbf P )
\nonumber
\\
&&
=( \mathbf I, \;  J  \; \mathbf P ).
\nonumber
 \end{eqnarray}
 Here we used that $( \mathbf I, \;  {\cal L } \; \mathbf X )=0$ for arbitrary vector $\mathbf X$.
 The normalised joint WTD for two waiting time  is
 \begin{eqnarray}
w_2(\tau_2,\tau_1) = 
\frac{( \mathbf I, \; J \; e^{ ({\cal L} -J ) \tau_2}   \; J \; e^{ ({\cal L} -J ) \tau_1} \; J  \; \mathbf P )}{( \mathbf I, \;  J  \; \mathbf P )}.
\label{wtd2-2}
\end{eqnarray}
The WTD $w_1(\tau)$ has the same normalisation (easy to show by computing the integral over $\tau$)\cite{kosov17-wtd}
 \begin{equation}
w_1(\tau) = \frac{  (\mathbf I, \; J \; e^{ ({\cal L} -J ) \tau} \; J \;  \mathbf P) }{  (\mathbf I, \;  J \;  \mathbf P)}. 
\label{wtd1}
\end{equation}
The formal derivation of WTDs (\ref{wtd2-2}) and (\ref{wtd1}) is presented in appendix A.
 
 Let us also check that these definitions (\ref{wtd2-2}) and (\ref{wtd1}) are consistent with each other. Integrating  $w_2$  over the second time yields
 \begin{eqnarray}
&&\int_0^{\infty} d \tau_2 \; w_2(\tau_2, \tau_1) 
\nonumber
\\
&&=-\frac{( \mathbf I, \; J \; ({\cal L} -J )^{-1}  \; J \; e^{ ({\cal L} -J ) \tau_1} \; J  \; \mathbf P )}{( \mathbf I, \;  J  \; \mathbf P )}
\nonumber
\\
&&=-\frac{( \mathbf I, \;  (J -{\cal L} + {\cal L})  \; ({\cal L} -J )^{-1}  \; J \; e^{ ({\cal L} -J ) \tau_1} \; J  \; \mathbf P )}{( \mathbf I, \;  J  \; \mathbf P )}
\nonumber
\\
&&=\frac{( \mathbf I, \; J \; e^{ ({\cal L} -J ) \tau_1} \; J  \; \mathbf P )}{( \mathbf I, \;  J  \; \mathbf P )} = w_1(\tau_1).
\nonumber
 \end{eqnarray}
Performing integration over the first waiting time (and using ${\cal L} \mathbf P=0$) gives
 \begin{eqnarray}
&&\int_0^{\infty} d \tau_1\; w_2(\tau_2, \tau_1) 
\nonumber
\\
&&=-\frac{( \mathbf I, \; J \; e^{ ({\cal L} -J ) \tau_2} \; J \; ({\cal L} -J )^{-1} \; J  \; \mathbf P )}{( \mathbf I, \;  J  \; \mathbf P )}
\nonumber
\\
&&=-\frac{( \mathbf I, \; J \; e^{ ({\cal L} -J ) \tau_2} \; J \; ({\cal L} -J )^{-1} \;  (J -{\cal L} + {\cal L})  \; \mathbf P )}{( \mathbf I, \;  J  \; \mathbf P )}
\nonumber
\\
&&=\frac{( \mathbf I, \; J \; e^{ ({\cal L} -J ) \tau_2} \; J  \; \mathbf P )}{( \mathbf I, \;  J  \; \mathbf P )} = w_1(\tau_2).
\nonumber
 \end{eqnarray}
Therefore,  our definitions for single and double time WTDs are consistent with each and have clear probabilistic meaning.

To compute higher-order expectation values and analyse the fluctuations, we introduce the cumulant-generating functions for the joint waiting time probability distribution
 \begin{eqnarray}
 K(x_1,x_2)= \int_0^\infty d \tau_1 \int_0^\infty  d \tau_2  \; e^{i x_1 \tau_1} e^{i x_2 \tau_2} w_2(\tau_1,\tau_2).
 \label{k}
 \end{eqnarray}
 Integrating the cumulant-generating function over $\tau_1$ and $\tau_2$, we get
 \begin{eqnarray}
\nonumber 
K(x_1,x_2) 
= \frac{  (\mathbf I, \; J \;  G(x_1)   \; J \; G(x_2)   \; J \;   \mathbf P) }{  (\mathbf I, \;  J \;  \mathbf P)},
\label{cum}
\end{eqnarray}
where 
\begin{equation}
G(x) = ({\cal L} -J+ix  )^{-1}.
\end{equation}
We obtain all possible higher order comulants differentiating $K(x_1, x_2)$ with respect to $x_1$ and $x_2$.

 \section{Results}
 
 \subsection{Equilibrium molecular vibrations}
 Master equation (\ref{me1},\ref{me2})  describes the non-equilibrium dynamics of molecular vibrations.
 Let us first consider the limit where the vibration is maintained in thermodynamic equilibrium at some temperature $T$, which is not necessarily the same  as the temperature of electrons in the leads.
 To implement this limit we use the following separable ansatz for the probabilities\cite{PhysRevB.69.245302}
 \begin{equation}
 P_{nq}(t) =P_n(t) \frac{e^{-q \omega/T}}{1-e^{-\omega/T}},
 \end{equation}
 which assumes that  the vibration maintains the equilibrium distribution at all time.
The master equation (\ref{me1},\ref{me2}) is reduced to
\begin{eqnarray*}
\frac{d}{dt} \left[\begin{array}{c}
 P_0\\
 P_1
\end{array}\right] & = & \left[\begin{array}{cc}
- \Gamma_{10} &  \Gamma_{01}\\
 \Gamma_{10} & - \Gamma_{01}
\end{array}\right]\left[\begin{array}{c}
P_0\\
P_1
\end{array}\right],
\label{rate}
\end{eqnarray*}
where the vibration averaged  rates are defined as
\begin{equation}
 \Gamma^\alpha_{mn} = \sum_{qq'} \Gamma^\alpha_{mq,nq'} \frac{e^{-q' \omega/T}}{1-e^{-\omega/T}}.
\end{equation}
Let us identify the quantum jump operator for the  electron tunneling  from the molecule to the drain electrode. We write this jump operator in matrix form and as a  dyadic product of two vectors
\begin{equation}
\label{jmd}
 {J}  = \left[\begin{array}{cc}
0 &  \Gamma_{01}^D \\
0 & 0
\end{array}\right]= 
\Gamma^D_{01}
 \left[\begin{array}{c}
1\\
0
\end{array}\right]\left[\begin{array}{cc}
0 &  1  \end{array}\right].
\end{equation}
Then straightforward vector algebra brings the WTD $w_1(\tau)$  (\ref{wtd1}) to the following form
 \begin{equation}
w_1(\tau) = \Gamma_{01}^D  
\left[\begin{array}{cc}
0 & 1 \end{array}\right]
e^{ ({\cal L} -J )\tau }
\left[\begin{array}{c}
1\\
0
\end{array}\right].
\end{equation} 
 For WTD $w_2(\tau_1, \tau_2)$  (\ref{wtd2-2}) we have
 \begin{widetext}
 \begin{equation}
w_2(\tau_2,\tau_2) =  \Gamma_{01}^D 
\left[\begin{array}{cc}
0& 1\end{array}\right]
e^{ ({\cal L} -J )\tau_2 }
\left[\begin{array}{c}
1\\
0
\end{array}\right]
 \Gamma_{01}^D 
\left[\begin{array}{cc}
0 & 1\end{array}\right]
e^{ ({\cal L} -J )\tau_1 }
\left[\begin{array}{c}
1\\
0
\end{array}\right]= w_1(\tau_2) w_1(\tau_1).
\end{equation} 
 \end{widetext}
We see that $w_2$  is always exactly factorised as a product of two independent $w_1$.
Therefore,  if the molecular vibration is held in thermal equilibrium, then the electronic distribution is always fully reset after each tunneling events and there is no correlation between subsequent  tunneling electrons.

 \subsection{Nonequilibrium molecular vibrations}
Let us now turn our attention to the case of fully nonequilibrium dynamics of molecular vibrations. We must rely on the numerical calculations to get answers in this situation. 
 
\begin{figure}[b]
\begin{center}
\includegraphics[width=1.05\columnwidth]{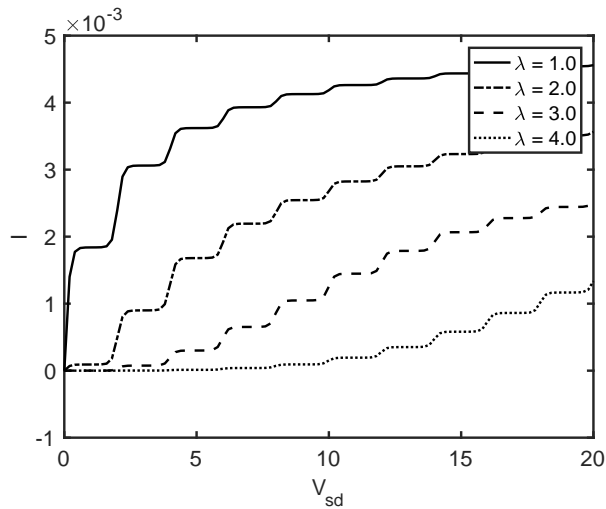}
\end{center}
	\caption{Current $I$ as a function of applied voltage  $V_{sd}$. Parameters used in calculations (all energy values are given in units of $\omega$): $\omega=1$,$\gamma_S = \gamma_D =0.01$, $ T =0.05$, 
$\epsilon =0$. Unit for  electric current is $\omega$ (or if we put $\hbar$ and $e$ back, it is  $ e \omega$) and  values of voltage bias $V_{sd}$ are given in $\omega$ (or $\hbar \omega/e$).}	
\label{IV}
\end{figure}

We first compute electric current as a function of the applied voltage bias $V_{sd}$. The voltage bias is enforced by shifting symmetrically the chemical potentials of the electrodes $ \mu_S = V_{sd}/2$ and $\mu_D = -V_{sd}/2 $. Fig.\ref{IV} shows the current-voltage characteristics. It has been studied in various details in many works before
\cite{galperin07,PhysRevB.83.115414,moletronics} and we show it here simply to serve as a reference - the characteristics steps in the current-voltage characteristic  will be  related to the  behaviour of the waiting time. The steps  in the current are  due to the resonant excitations of the  molecular vibrations by  inelastic tunneling of electrons. The steps are observed when the voltage passes through an integer multiple  of the vibration energy.   We also observe the current suppression in the strong electron-vibration coupling regime due to Franck-Condon blockade.\cite{PhysRevB.73.155306}

\begin{figure}[b]
\begin{center}
\includegraphics[width=1.05\columnwidth]{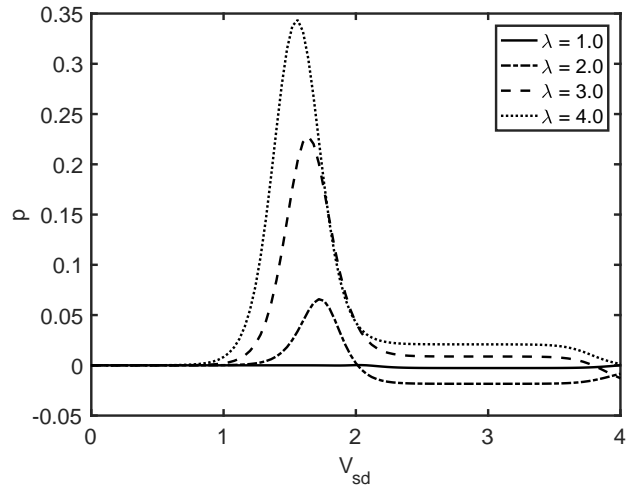}
\end{center}
	\caption{Pearson correlation coefficient between two subsequent tunneling events, $ p= (\langle  \tau_1 \tau_2 \rangle - \langle \tau \rangle^2)/(  \langle   \tau^2 \rangle- \langle   \tau \rangle^2 ) $ as a function of applied voltage  $V_{sd}$ computed for different strengths of electron-vibration coupling. Parameters used in calculations (all energy values are given in units of $\omega$): $\omega=1$, $\gamma_S = \gamma_D =0.01$, $ T =0.05$, 
$\epsilon =0$.  The voltage bias $V_{sd}$ is given in $\omega$.}	
\label{fig-pearson}
\end{figure}

\begin{figure}[t]
\begin{center}
\includegraphics[width=1.05\columnwidth]{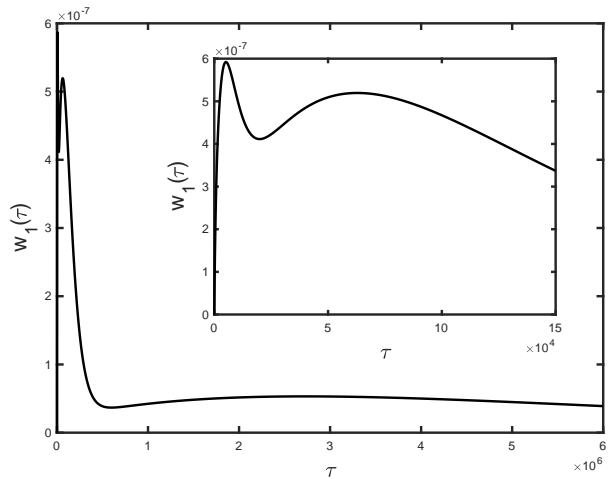}
\end{center}
	\caption{Waiting time distribution $w_1(\tau)$ for $\lambda=4$ and $V_{sd} =1.56 \omega$. Parameters used in calculations (all energy values are given in units of $\omega$): 
	$\omega=1$, $\gamma_S = \gamma_D =0.01$, $ T =0.05$, 
$\epsilon =0$. Insert figure zooms the WTD in the short waiting time. Unit for time is $1/\omega$.}	
\label{fig-wtd}
\end{figure}

\begin{figure}[t]
\begin{center}
\includegraphics[width=1.05\columnwidth]{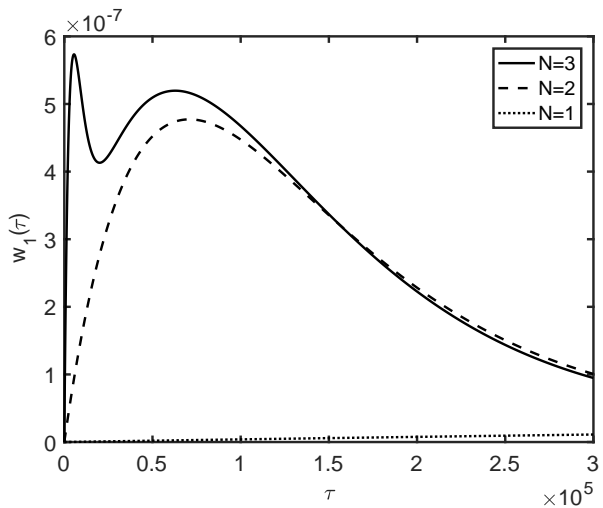}
\end{center}
	\caption{Waiting time distribution $w_1(\tau)$ for $\lambda=4$ and $V_{sd} =1.56 \omega$ computed for different numbers of vibrational occupation cutoffs. Parameters used in calculations (all energy values are given in units of $\omega$): $\omega=1$, $\gamma_S = \gamma_D =0.01$, $ T =0.05$, 
$\epsilon =0$.  Unit for time is $1/\omega$.}	
\label{fig-wtdzoom}
\end{figure}

 Correlations between two subsequent tunneling events will be measured using Pearson correlation coefficient
 \begin{equation}
 p=\frac{\langle  \tau_1 \tau_2 \rangle - \langle \tau \rangle^2 }{  \langle   \tau^2 \rangle- \langle   \tau \rangle^2 }.
 \label{pearson}
 \end{equation}
Integrals in the Pearson coefficient  are computed using cumulant generating function (\ref{cum}):
For the correlations we have
\begin{eqnarray}
&& \langle  \tau_2 \tau_1 \rangle 
= 
\int_0^{\infty} d \tau_1 \int_0^{\infty} d \tau_2 \; \tau_1 \tau_2 w_2(\tau_2, \tau_1)
\nonumber
\\
&& = 
\frac{  (\mathbf I, \; J \;  G(0)^2     \; J \; G(0)^2    \; J \;   \mathbf P) }{  (\mathbf I, \;  J \;  \mathbf P)}
\end{eqnarray}
and  moments of single waiting time are
\begin{equation}
 \langle \tau^n \rangle =
\int_0^{\infty} d \tau_1  \; \tau^n w_1(\tau) 
 =  n! (-1)^{n+1} 
\frac{  (\mathbf I, \; J \; G(0)^{n+1}   \; J \;   \mathbf P) }{  (\mathbf I, \;  J \;  \mathbf P)}.
\end{equation}
The Pearson correlation coefficient  $p$ is widely used in statistics as a measure of correlations between two stochastic variables. It varies between $-1$ and $+1$: $p=0$ means that there is no correlations, $p>0$ indicates positive  correlations,
and $p<0$ suggests the  variables are anticorrelated (negative correlations). In our case positive correlations mean that 
if the first waiting time increases/decreases then the waiting time for the second electron also increases/decreases. The negative correlations indicate that 
if the first electron waits longer then the waiting time for the second electron decreases  (and vice versa).

Fig.2 shows the Pearson correlation coefficient between successive electron tunneling events (\ref{pearson}) 
computed  as a function of the applied voltage bias for different values of electron-vibration coupling. 
If there is electron-vibration coupling, then the waiting time for successive electron tunneling are not correlated. The correlation 
does not appear in the weak and moderate electron-vibration coupling regimes (for $\lambda <1$). Only when the electron-vibrational interaction becomes strong $\lambda>1$, the statistical correlations between electronic tunneling events start to emerge. 

We focus on the strong coupling regime  ($\lambda=2 $, $\lambda= 3 $ and $\lambda=4 $ in Fig.2), this is where the voltage dependence of the Pearson correlation coefficient is very interesting. The tunneling events are positively correlated in the narrow voltage window $\omega \le V_{sd} \le 2  \omega$ (the negative correlations for $\lambda=2$  all have $p < 0.02$ -- they are statistically negligible). Comparing Fig.1 and Fig.2, we see positive correlations belong to the Franck-Condon blockade regime where the electric current is small and, therefore, the tunneling events are extremely rare.

Let us understand why the electron tunneling events become suddenly positively correlated. First, we compute WTD $w_1(\tau_1)$ for $\lambda=4$ and $V_{sd} =1.56 \; \omega$. The dependence of $w_1(\tau_1)$ on the waiting time is shown in Fig.\ref{fig-wtd}. The
distribution has a large double peaked spike for the fast electrons (the first narrow peak is barely seen on the main figure since it is very close to the axis) - the structure of the spike is zoomed in the insert of
Fig.\ref{fig-wtd}. To understand the origin of this behaviour of WTD, we compute waiting time distribution for various cutoffs  $N$ for the number of the vibrational quanta included in the calculations (Fig.4). For $N=1$ (it means that only states with vibrational quantum numbers $q=0,1$ are included in the calculations), the short time peaks disappear completely. 
For $N=2$ (vibrational quantum numbers $q=0,1,2$ are included), the short time  WTD spike consists only of the one broad peak.  For $N=3$ (vibrational quantum numbers $q=0,1,2,3$ are included), the short time WTD spike forms its final two peak structure. Increasing $N$ further does not affect the short time behaviour of WTD at the considered voltage.

We are now in the position to explain the physical reasons behind the appearance of the positive temporal correlations. The mode of the distribution is the first early time peak - it gives us the waiting time for the electrons most often observed in the transport. These fast travelling electrons are responsible for the emergence of the correlations between successive electron tunneling events. We focus on the regime where we observe positive correlations ($\lambda =4 $ and $V_{sd}=1.56 \; \omega$). After the first quantum jump the first $q=1$ and second  $q=2$  vibrational states become populated in the density matrix (\ref{jp}). Then, if the first waiting time $\tau_1$ is around the mode of the waiting time  distribution $w_1(\tau_1)$ (Fig.\ref{fig-wtd}),  the vibrational state $q=3$ is predominantly populated after the waiting time $\tau_1$(see the appendix for  details).
If we look at the absolute values of the  first 4 components of the Franck-Condon factor $|X_{qq'}|, \; q,q'=0,..,3$ computed for $\lambda=4 $
\begin{equation}
|X|=\left(
\begin{array}{cccc}
0.0003 &  0.0013 & 0.0038 & 0.0088 \\
0.0013 & 0.0050 & 0.0133 & 0.0285 \\
0.0038 &0.0133 & 0.0325 & 0.0643 \\
 0.0088 & 0.0285 &0.0643 & 0.1160
\end{array}
 \right)
\end{equation}
we see that it creates the "shortcut" opening the elastic transport channel  through the vibrational state $q=3$. This elastic channel opens only for a short time during the first narrow peak in the WTD $w_1(\tau)$ (Fig.\ref{fig-wtd}). The higher states $q=4,5,...$ have even larger diagonal Franck-Condon factor but they are not populated on the short time scale at this voltage range. At the higher voltages many vibrational states are populated already in the steady state density matrix, but it does not lead to the correlations since too many elastic channels are available anyway irrespective to the previous tunneling electron.

These temporal correlations between tunneling times should not be confused with "avalanche" electron transport phenomena in molecular junctions with strong electron-vibrational coupling.\cite{fcblockade05,PhysRevB.73.155306,doi:10.1021/acs.nanolett.5b03434} The "avalanche" transport of electrons is  observed at intermediate voltage range $3 \omega <V_{sd}<6 \omega$ just after the Franck-Condon plato in I-V characteristics and it does not involve the correlations between different waiting times.\cite{fcblockade05}  The avalanche electron transport is simply manifestation of the fact that the mode of WTD is much smaller than the average time given by the same distribution.

\section{Conclusions}

In this paper, we developed the theory for computing  joint waiting time distribution for electron transport through a molecular junction with strong electron-vibrational interaction.
The molecule is modelled by one molecular orbital coupled with a single localised vibration. We treat electron-vibration interaction exactly and molecule-electrode coupling  within the Born-Markov approximation. Using this master equation we computed joint waiting time probability distribution and studied it across various transport regimes to understand the emergence and disappearance of the correlations between successive electron tunneling events.

Our main observations are summarised below:

\begin{itemize}
\item 
There are no temporal correlations between subsequent electron tunneling events (the distribution function is completely renewed  after each electron tunneling) 
\begin{enumerate}
\item[(a)]
for small voltage bias ($<\omega$) and for voltages greater than $2 \omega$;
\item[(b)]
irrespective of voltage bias if electron-vibration coupling $\lambda < 1$;

\item[(c)]
 irrespective to voltage and electron-vibration coupling strength if the vibration is maintained in thermodynamic equilibrium.
\end{enumerate}

\item
The temporal  correlations between subsequent electron tunnelings emerge,  if $\omega< V_{sd}< 2 \omega $ and $\lambda>1$.
 The tunneling events become positively correlated which means that the second electron can sense the waiting time for the first electron and if the first electron was fast, then the second electron also would like to be transferred quickly (and opposite, slow to slow correlations are also possible).  The observed results are robust, they do not require any special tuning of the parameters other than  the physically reasonable choices of applied voltage and  electron-vibration coupling.

\item
The physical origin of positively correlated waiting times is the following.
After the initial electron transfer to the drain electrode the ground and the first excited    vibrational states are predominantly populated. Next, if the first waiting time  happens to be around the modal time,  then the  electron has an opportunity to excite the molecule in $q=3$ state via tunneling through inelastic $q=0$ to $q=3$ or  $q=1$ to $q=3$ channels.
The excitation of $q=3$ vibrational state
  creates a  "shortcut" between the source and drain electrodes via  the elastic channel  with large Franck-Condon factor which can be utilised by the next electron. In other words, if the the first waiting time is the modal time (very short), then the tunneling electron opens $q=3$ elastic channel for swift  transfer for the next electron (positive correlations).

\end{itemize}

\appendix

\section{Waiting time distributions for electron detection events}
In this appendix, we define  WTDs (\ref{wtd1}) and (\ref{wtd2-2}) using methods of quantum measurement theory.
Our derivations  follow  the theory originally  developed  in quantum optics to study single photon counting statistics  \cite{Srinivas2010,Zoller1987,nonrenewal-budini} and extended to quantum  transport by Brandes.\cite{brandes08}

The Liouvillian in the master equation (\ref{master-eq}) is decomposed as
\begin{equation}
\dot {\mathbf P}(t) =( {\cal L}_0 + J ) \mathbf P(t),
\end{equation}
where  $J$ is quantum jump operator (\ref{jump}) and ${\cal L}_0 = {\cal L} -J $ generates the evolution of the molecular junction without transferring electrons to the drain electrode.  
This differential equation is converted to the integral equation
\begin{equation}
\mathbf P(t) =e^{{\cal L}_0 t}   \mathbf P(0) + \int_0^t dt_1 e^{{\cal L}_0 (t-t_1)}  J \mathbf P(t_1),
\label{master-int}
\end{equation}
which then is resolved by iterations
\begin{align}
\label{master-int1}
&\mathbf P(t) =e^{{\cal L}_0 t}   \mathbf P(0) +  \int_0^t d t_1  e^{{\cal L}_0 (t-t_1)}  J    e^{{\cal L}_0 t_1} \mathbf P(0)
 \\
&+  \int_0^t d t_1   \int_0^{t_1} d t_2 e^{{\cal L}_0 (t-t_1)}  J    
e^{{\cal L}_0 (t_1-t_2)} J e^{{\cal L}_0 t_2} \mathbf P(0) + ....
\nonumber
\end{align}
We introduce operator $M$ - it describes detection of electron transfer from the molecule to the drain electrode (quantum measurement operator).
If $\mathbf P(t)$ is the probability vector before the measurement, then after the electron detection it becomes\cite{Petruccione}
\begin{equation}
M \mathbf P(t) = \frac{ J \mathbf P(t)}{(\mathbf I, J \mathbf P(t))}.
\label{m}
\end{equation}

We choose the initial probability vector as
\begin{equation}
\mathbf P(0) = M \mathbf P,
\end{equation}
where $\mathbf P$ is the steady state probability vector defined in (\ref{ss}). This choice of the initial state means  that we detect the electron transfer to the drain electrode at time $t=0$ in the steady state regime and then we start to monitor the system:
\begin{align}
\label{master-int2}
&\mathbf P(t) =e^{{\cal L}_0 t}   M \mathbf P +  \int_0^t d t_1  e^{{\cal L}_0 (t-t_1)}  J    e^{{\cal L}_0 t_1} M \mathbf P
 \\
&+  \int_0^t d t_1   \int_0^{t_1} d t_2 e^{{\cal L}_0 (t-t_1)}  J    
e^{{\cal L}_0 (t_1-t_2)} J e^{{\cal L}_0 t_2} M \mathbf P + ....
\nonumber
\end{align}
Using electron detection operator (\ref{m}),
we rewrite (\ref{master-int1}) in a form which elucidate the probabilistic meanings of its terms:\cite{Srinivas2010,Zoller1987,nonrenewal-budini}
\begin{widetext}
\begin{align}
\label{master-int3}
&\mathbf P(t) =e^{{\cal L}_0 t}   M \mathbf P 
+  \int_0^t d t_1  \frac{(\mathbf I, J e^{{\cal L}_0 t_1} J \mathbf P)}{(\mathbf I, J \mathbf P)} \;  e^{{\cal L}_0 (t-t_1)}   M   e^{{\cal L}_0 t_1} M \mathbf P
 \\
&
+  \int_0^t d t_1   \int_0^{t_1} d t_2   \frac{(\mathbf I, J   e^{{\cal L}_0 (t_1-t_2)} J e^{{\cal L}_0 t_1} J \mathbf P)}{(\mathbf I, J \mathbf P)} \;
e^{{\cal L}_0 (t-t_1)}  M   e^{{\cal L}_0 (t_1-t_2)} M e^{{\cal L}_0 t_2} M \mathbf P + ....
\nonumber
\end{align}
\end{widetext}

Let us discuss (\ref{master-int3}). We begin with first term, $ e^{{\cal L}_0 t}  M \mathbf P$, it is the contribution to the probability vector  from all measurements  where no electron transfer to the drain electron to occur up to time $t$ after the initial detection at time $t=0$.  

The first integral term in this equation can be read as the following: an electron is detected in the drain electrode at time $t=0$ (due to the presence of $M \mathbf P$), then no detection of electron is observed up to time $t_1$ (due to presence of  the "idle" evolution operator $e^{{\cal L}_0 t_1}$),
then the detection of the second electron occurs at time $t_1$,  and then the system "idle" without electron transfer to the drain electrode up to time $t$.
 Therefore, the waiting prefactor $(\mathbf I, J e^{{\cal L}_0 t_1} J \mathbf P)/(\mathbf I, J \mathbf P)$ must be understood as the probability of observing this process. 
 This waiting factor is exactly our  expression (\ref{wtd1}) for normalized WTD $w_1(t_1)$.   The analysis of the second integral term follows  exactly the same lines and the waiting prefactor $(\mathbf I, J   e^{{\cal L}_0 (t_1-t_2)} J e^{{\cal L}_0 t_1} J \mathbf P)/(\mathbf I, J \mathbf P)$ is interpreted as WTD $w_2(t_1-t_2,t_2)$ (\ref{wtd2-2}).

\section{Probability vectors after quantum jumps}
We show in this appendix the probability vectors (normalised) after the quantum jumps associated with electron tunneling to the drain electrode. All probability vectors are computed at the voltage, which corresponds to the maximum Pearson correlation coefficient. The electron-vibration coupling is  $\lambda =4$.

We show only first  $q=0,1,2,3$ states from the full probability vector (\ref{P}). Steady state probability and the probability vector (normalized) after the quantum jump are
\begin{eqnarray}
\mathbf P=\left(
\begin{array}{c} 
   0.5 \\
   0.5 \\
   0.0 \\
   0.0\\
   0.0\\
   0.0 \\
   0.0 \\
   0.0
\end{array}
 \right), \;\; \;\;  J \mathbf P=\left(
\begin{array}{c} 
   0.4 \\ 
   0.0 \\
   0.5 \\
   0.0 \\
   0.1 \\
   0.0 \\
   0.0 \\
   0.0 
\end{array}
 \right).
 \label{jp}
\end{eqnarray}
Now we compare  $J e^{ {\cal L}_0 \tau} J \mathbf P$  for different waiting times $\tau=0.05\times10^5$ (corresponds to the first peak in WTD shown on insert plot in Fig.{\ref{fig-wtd}), $\tau=10^5$ (corresponds to the second peak in the WTD shown in insert plot in Fig.{\ref{fig-wtd} and $\tau=5\times10^6$ (tail of the WTD):
\begin{eqnarray}
\left(
\begin{array}{c} 
   0.0  \\                   
   0.0 \\
   0.1  \\                
   0.0 \\
   0.3  \\
   0.0 \\
   0.6 \\
   0.0
\end{array}
 \right),
 \;\;\;\;
\left(
\begin{array}{c} 
   0.0 \\
   0.0 \\
   0.2 \\
   0.0 \\
   0.8 \\
   0.0 \\
   0.0 \\
   0.0
\end{array}
 \right), \;\; \;\;  
\left(
\begin{array}{c} 
   0.1 \\
   0.0 \\
   0.8 \\
   0.0 \\
   0.1 \\
   0.0 \\
   0.0 \\
   0.0
\end{array}
 \right).
\end{eqnarray}
From these probability vectors, we see that electron tunneling during the first peak waiting time populates predominantly $q=3$ vibrational state;
electron tunneling during the second peak waiting time populates predominantly $q=2$ vibrational state, and then waiting long time to tunnel brings the probability close to the  initial $J\mathbf P$ vector.

\clearpage

\begin{thebibliography}{41}
\expandafter\ifx\csname natexlab\endcsname\relax\def\natexlab#1{#1}\fi
\expandafter\ifx\csname bibnamefont\endcsname\relax
  \def\bibnamefont#1{#1}\fi
\expandafter\ifx\csname bibfnamefont\endcsname\relax
  \def\bibfnamefont#1{#1}\fi
\expandafter\ifx\csname citenamefont\endcsname\relax
  \def\citenamefont#1{#1}\fi
\expandafter\ifx\csname url\endcsname\relax
  \def\url#1{\texttt{#1}}\fi
\expandafter\ifx\csname urlprefix\endcsname\relax\def\urlprefix{URL }\fi
\providecommand{\bibinfo}[2]{#2}
\providecommand{\eprint}[2][]{\url{#2}}

\bibitem[{\citenamefont{Nazarov and Blanter}(2009)}]{nazarov-book}
\bibinfo{author}{\bibfnamefont{Y.~V.} \bibnamefont{Nazarov}} \bibnamefont{and}
  \bibinfo{author}{\bibfnamefont{Y.~M.} \bibnamefont{Blanter}},
  \emph{\bibinfo{title}{Quantum Transport: Introduction to Nanoscience}}
  (\bibinfo{publisher}{Cambridge University Press}, \bibinfo{year}{2009}).

\bibitem[{\citenamefont{Thomas and Flindt}(2013)}]{flindt13}
\bibinfo{author}{\bibfnamefont{K.~H.} \bibnamefont{Thomas}} \bibnamefont{and}
  \bibinfo{author}{\bibfnamefont{C.}~\bibnamefont{Flindt}},
  \bibinfo{journal}{Phys. Rev. B} \textbf{\bibinfo{volume}{87}},
  \bibinfo{pages}{121405} (\bibinfo{year}{2013}).

\bibitem[{\citenamefont{Sothmann}(2014)}]{sothmann14}
\bibinfo{author}{\bibfnamefont{B.}~\bibnamefont{Sothmann}},
  \bibinfo{journal}{Phys. Rev. B} \textbf{\bibinfo{volume}{90}},
  \bibinfo{pages}{155315} (\bibinfo{year}{2014}).

\bibitem[{\citenamefont{Thomas and Flindt}(2014)}]{flindt14}
\bibinfo{author}{\bibfnamefont{K.~H.} \bibnamefont{Thomas}} \bibnamefont{and}
  \bibinfo{author}{\bibfnamefont{C.}~\bibnamefont{Flindt}},
  \bibinfo{journal}{Phys. Rev. B} \textbf{\bibinfo{volume}{89}},
  \bibinfo{pages}{245420} (\bibinfo{year}{2014}).

\bibitem[{\citenamefont{Potanina and Flindt}(2017)}]{flindt17}
\bibinfo{author}{\bibfnamefont{E.}~\bibnamefont{Potanina}} \bibnamefont{and}
  \bibinfo{author}{\bibfnamefont{C.}~\bibnamefont{Flindt}},
  \bibinfo{journal}{Phys. Rev. B} \textbf{\bibinfo{volume}{96}},
  \bibinfo{pages}{045420} (\bibinfo{year}{2017}).

\bibitem[{\citenamefont{Tang et~al.}(2014)\citenamefont{Tang, Xu, and
  Wang}}]{wtd-transient}
\bibinfo{author}{\bibfnamefont{G.-M.} \bibnamefont{Tang}},
  \bibinfo{author}{\bibfnamefont{F.}~\bibnamefont{Xu}}, \bibnamefont{and}
  \bibinfo{author}{\bibfnamefont{J.}~\bibnamefont{Wang}},
  \bibinfo{journal}{Phys. Rev. B} \textbf{\bibinfo{volume}{89}},
  \bibinfo{pages}{205310} (\bibinfo{year}{2014}).

\bibitem[{\citenamefont{Seoane~Souto et~al.}(2015)\citenamefont{Seoane~Souto,
  Avriller, Monreal, Mart\'{\i}n-Rodero, and Levy~Yeyati}}]{PhysRevB.92.125435}
\bibinfo{author}{\bibfnamefont{R.}~\bibnamefont{Seoane~Souto}},
  \bibinfo{author}{\bibfnamefont{R.}~\bibnamefont{Avriller}},
  \bibinfo{author}{\bibfnamefont{R.~C.} \bibnamefont{Monreal}},
  \bibinfo{author}{\bibfnamefont{A.}~\bibnamefont{Mart\'{\i}n-Rodero}},
  \bibnamefont{and}
  \bibinfo{author}{\bibfnamefont{A.}~\bibnamefont{Levy~Yeyati}},
  \bibinfo{journal}{Phys. Rev. B} \textbf{\bibinfo{volume}{92}},
  \bibinfo{pages}{125435} (\bibinfo{year}{2015}).

\bibitem[{\citenamefont{Goswami and Harbola}(2015)}]{harbola15}
\bibinfo{author}{\bibfnamefont{H.~P.} \bibnamefont{Goswami}} \bibnamefont{and}
  \bibinfo{author}{\bibfnamefont{U.}~\bibnamefont{Harbola}},
  \bibinfo{journal}{J. Chem. Phys.} \textbf{\bibinfo{volume}{142}}
  (\bibinfo{year}{2015}).

\bibitem[{\citenamefont{Rudge and Kosov}(2016{\natexlab{a}})}]{rudge16a}
\bibinfo{author}{\bibfnamefont{S.~L.} \bibnamefont{Rudge}} \bibnamefont{and}
  \bibinfo{author}{\bibfnamefont{D.~S.} \bibnamefont{Kosov}},
  \bibinfo{journal}{J. Chem. Phys.} \textbf{\bibinfo{volume}{144}},
  \bibinfo{eid}{124105} (\bibinfo{year}{2016}{\natexlab{a}}).

\bibitem[{\citenamefont{Rudge and Kosov}(2016{\natexlab{b}})}]{rudge16b}
\bibinfo{author}{\bibfnamefont{S.~L.} \bibnamefont{Rudge}} \bibnamefont{and}
  \bibinfo{author}{\bibfnamefont{D.~S.} \bibnamefont{Kosov}},
  \bibinfo{journal}{Phys. Rev. E} \textbf{\bibinfo{volume}{94}},
  \bibinfo{pages}{042134} (\bibinfo{year}{2016}{\natexlab{b}}).

\bibitem[{\citenamefont{Kosov}(2017)}]{kosov17-wtd}
\bibinfo{author}{\bibfnamefont{D.~S.} \bibnamefont{Kosov}},
  \bibinfo{journal}{J. Chem. Phys.} \textbf{\bibinfo{volume}{146}},
  \bibinfo{pages}{074102} (\bibinfo{year}{2017}).

\bibitem[{\citenamefont{Brandes}(2008)}]{brandes08}
\bibinfo{author}{\bibfnamefont{T.}~\bibnamefont{Brandes}},
  \bibinfo{journal}{Ann. Phys. (Berlin)} \textbf{\bibinfo{volume}{17}},
  \bibinfo{pages}{477} (\bibinfo{year}{2008}).

\bibitem[{\citenamefont{Albert et~al.}(2012)\citenamefont{Albert, Haack,
  Flindt, and B\"uttiker}}]{buttiker12}
\bibinfo{author}{\bibfnamefont{M.}~\bibnamefont{Albert}},
  \bibinfo{author}{\bibfnamefont{G.}~\bibnamefont{Haack}},
  \bibinfo{author}{\bibfnamefont{C.}~\bibnamefont{Flindt}}, \bibnamefont{and}
  \bibinfo{author}{\bibfnamefont{M.}~\bibnamefont{B\"uttiker}},
  \bibinfo{journal}{Phys. Rev. Lett.} \textbf{\bibinfo{volume}{108}},
  \bibinfo{pages}{186806} (\bibinfo{year}{2012}).

\bibitem[{\citenamefont{Dasenbrook et~al.}(2015)\citenamefont{Dasenbrook,
  Hofer, and Flindt}}]{flindt15}
\bibinfo{author}{\bibfnamefont{D.}~\bibnamefont{Dasenbrook}},
  \bibinfo{author}{\bibfnamefont{P.~P.} \bibnamefont{Hofer}}, \bibnamefont{and}
  \bibinfo{author}{\bibfnamefont{C.}~\bibnamefont{Flindt}},
  \bibinfo{journal}{Phys. Rev. B} \textbf{\bibinfo{volume}{91}},
  \bibinfo{pages}{195420} (\bibinfo{year}{2015}).

\bibitem[{\citenamefont{Koch and von Oppen}(2005)}]{fcblockade05}
\bibinfo{author}{\bibfnamefont{J.}~\bibnamefont{Koch}} \bibnamefont{and}
  \bibinfo{author}{\bibfnamefont{F.}~\bibnamefont{von Oppen}},
  \bibinfo{journal}{Phys. Rev. Lett.} \textbf{\bibinfo{volume}{94}},
  \bibinfo{pages}{206804} (\bibinfo{year}{2005}).

\bibitem[{\citenamefont{Koch et~al.}(2006)\citenamefont{Koch, Semmelhack, von
  Oppen, and Nitzan}}]{PhysRevB.73.155306}
\bibinfo{author}{\bibfnamefont{J.}~\bibnamefont{Koch}},
  \bibinfo{author}{\bibfnamefont{M.}~\bibnamefont{Semmelhack}},
  \bibinfo{author}{\bibfnamefont{F.}~\bibnamefont{von Oppen}},
  \bibnamefont{and} \bibinfo{author}{\bibfnamefont{A.}~\bibnamefont{Nitzan}},
  \bibinfo{journal}{Phys. Rev. B} \textbf{\bibinfo{volume}{73}},
  \bibinfo{pages}{155306} (\bibinfo{year}{2006}).

\bibitem[{\citenamefont{Lau et~al.}(2016)\citenamefont{Lau, Sadeghi, Rogers,
  Sangtarash, Dallas, Porfyrakis, Warner, Lambert, Briggs, and
  Mol}}]{doi:10.1021/acs.nanolett.5b03434}
\bibinfo{author}{\bibfnamefont{C.~S.} \bibnamefont{Lau}},
  \bibinfo{author}{\bibfnamefont{H.}~\bibnamefont{Sadeghi}},
  \bibinfo{author}{\bibfnamefont{G.}~\bibnamefont{Rogers}},
  \bibinfo{author}{\bibfnamefont{S.}~\bibnamefont{Sangtarash}},
  \bibinfo{author}{\bibfnamefont{P.}~\bibnamefont{Dallas}},
  \bibinfo{author}{\bibfnamefont{K.}~\bibnamefont{Porfyrakis}},
  \bibinfo{author}{\bibfnamefont{J.}~\bibnamefont{Warner}},
  \bibinfo{author}{\bibfnamefont{C.~J.} \bibnamefont{Lambert}},
  \bibinfo{author}{\bibfnamefont{G.~A.~D.} \bibnamefont{Briggs}},
  \bibnamefont{and} \bibinfo{author}{\bibfnamefont{J.~A.} \bibnamefont{Mol}},
  \bibinfo{journal}{Nano Letters} \textbf{\bibinfo{volume}{16}},
  \bibinfo{pages}{170} (\bibinfo{year}{2016}).

\bibitem[{\citenamefont{H{\"a}rtle and Thoss}(2011)}]{PhysRevB.83.115414}
\bibinfo{author}{\bibfnamefont{R.}~\bibnamefont{H{\"a}rtle}} \bibnamefont{and}
  \bibinfo{author}{\bibfnamefont{M.}~\bibnamefont{Thoss}},
  \bibinfo{journal}{Phys. Rev. B} \textbf{\bibinfo{volume}{83}},
  \bibinfo{pages}{115414} (\bibinfo{year}{2011}).

\bibitem[{\citenamefont{Kuznetsov}(2007)}]{kuznetsov-ndr}
\bibinfo{author}{\bibfnamefont{A.~M.} \bibnamefont{Kuznetsov}},
  \bibinfo{journal}{J. Chem. Phys.} \textbf{\bibinfo{volume}{127}},
  \bibinfo{pages}{084710} (\bibinfo{year}{2007}).

\bibitem[{\citenamefont{Galperin et~al.}(2005)\citenamefont{Galperin, Ratner,
  and Nitzan}}]{galperin05}
\bibinfo{author}{\bibfnamefont{M.}~\bibnamefont{Galperin}},
  \bibinfo{author}{\bibfnamefont{M.~A.} \bibnamefont{Ratner}},
  \bibnamefont{and} \bibinfo{author}{\bibfnamefont{A.}~\bibnamefont{Nitzan}},
  \bibinfo{journal}{Nano Letters} \textbf{\bibinfo{volume}{5}},
  \bibinfo{pages}{125} (\bibinfo{year}{2005}).

\bibitem[{\citenamefont{Zazunov et~al.}(2006)\citenamefont{Zazunov, Feinberg,
  and Martin}}]{zazunov06}
\bibinfo{author}{\bibfnamefont{A.}~\bibnamefont{Zazunov}},
  \bibinfo{author}{\bibfnamefont{D.}~\bibnamefont{Feinberg}}, \bibnamefont{and}
  \bibinfo{author}{\bibfnamefont{T.}~\bibnamefont{Martin}},
  \bibinfo{journal}{Phys. Rev. B} \textbf{\bibinfo{volume}{73}},
  \bibinfo{pages}{115405} (\bibinfo{year}{2006}).

\bibitem[{\citenamefont{Dzhioev and Kosov}(2012)}]{solvent12}
\bibinfo{author}{\bibfnamefont{A.~A.} \bibnamefont{Dzhioev}} \bibnamefont{and}
  \bibinfo{author}{\bibfnamefont{D.~S.} \bibnamefont{Kosov}},
  \bibinfo{journal}{Phys. Rev. B} \textbf{\bibinfo{volume}{85}},
  \bibinfo{pages}{033408} (\bibinfo{year}{2012}).

\bibitem[{\citenamefont{Dzhioev and Kosov}(2011)}]{dzhioev11}
\bibinfo{author}{\bibfnamefont{A.~A.} \bibnamefont{Dzhioev}} \bibnamefont{and}
  \bibinfo{author}{\bibfnamefont{D.~S.} \bibnamefont{Kosov}},
  \bibinfo{journal}{J. Chem. Phys.} \textbf{\bibinfo{volume}{135}},
  \bibinfo{eid}{074701} (\bibinfo{year}{2011}).

\bibitem[{\citenamefont{Thomas et~al.}(2012)\citenamefont{Thomas, Karzig,
  Kusminskiy, Zar\'and, and von Oppen}}]{PhysRevB.86.195419}
\bibinfo{author}{\bibfnamefont{M.}~\bibnamefont{Thomas}},
  \bibinfo{author}{\bibfnamefont{T.}~\bibnamefont{Karzig}},
  \bibinfo{author}{\bibfnamefont{S.~V.} \bibnamefont{Kusminskiy}},
  \bibinfo{author}{\bibfnamefont{G.}~\bibnamefont{Zar\'and}}, \bibnamefont{and}
  \bibinfo{author}{\bibfnamefont{F.}~\bibnamefont{von Oppen}},
  \bibinfo{journal}{Phys. Rev. B} \textbf{\bibinfo{volume}{86}},
  \bibinfo{pages}{195419} (\bibinfo{year}{2012}).

\bibitem[{\citenamefont{Dzhioev et~al.}(2013)\citenamefont{Dzhioev, Kosov, and
  von Oppen}}]{catalysis12}
\bibinfo{author}{\bibfnamefont{A.~A.} \bibnamefont{Dzhioev}},
  \bibinfo{author}{\bibfnamefont{D.~S.} \bibnamefont{Kosov}}, \bibnamefont{and}
  \bibinfo{author}{\bibfnamefont{F.}~\bibnamefont{von Oppen}},
  \bibinfo{journal}{J. Chem. Phys.} \textbf{\bibinfo{volume}{138}},
  \bibinfo{eid}{134103} (\bibinfo{year}{2013}).

\bibitem[{\citenamefont{Galperin et~al.}(2009)\citenamefont{Galperin, Saito,
  Balatsky, and Nitzan}}]{galperin09}
\bibinfo{author}{\bibfnamefont{M.}~\bibnamefont{Galperin}},
  \bibinfo{author}{\bibfnamefont{K.}~\bibnamefont{Saito}},
  \bibinfo{author}{\bibfnamefont{A.~V.} \bibnamefont{Balatsky}},
  \bibnamefont{and} \bibinfo{author}{\bibfnamefont{A.}~\bibnamefont{Nitzan}},
  \bibinfo{journal}{Phys. Rev. B} \textbf{\bibinfo{volume}{80}},
  \bibinfo{pages}{115427} (\bibinfo{year}{2009}).

\bibitem[{\citenamefont{Ioffe et~al.}(2008)\citenamefont{Ioffe, Shamai, Ophir,
  Noy, Yutsis, Kfir, Cheshnovsky, and Selzer}}]{Ioffe:2008aa}
\bibinfo{author}{\bibfnamefont{Z.}~\bibnamefont{Ioffe}},
  \bibinfo{author}{\bibfnamefont{T.}~\bibnamefont{Shamai}},
  \bibinfo{author}{\bibfnamefont{A.}~\bibnamefont{Ophir}},
  \bibinfo{author}{\bibfnamefont{G.}~\bibnamefont{Noy}},
  \bibinfo{author}{\bibfnamefont{I.}~\bibnamefont{Yutsis}},
  \bibinfo{author}{\bibfnamefont{K.}~\bibnamefont{Kfir}},
  \bibinfo{author}{\bibfnamefont{O.}~\bibnamefont{Cheshnovsky}},
  \bibnamefont{and} \bibinfo{author}{\bibfnamefont{Y.}~\bibnamefont{Selzer}},
  \bibinfo{journal}{Nat Nano} \textbf{\bibinfo{volume}{3}},
  \bibinfo{pages}{727} (\bibinfo{year}{2008}).

\bibitem[{\citenamefont{Ptaszy\ifmmode~\acute{n}\else
  \'{n}\fi{}ski}(2017)}]{PhysRevB.95.045306}
\bibinfo{author}{\bibfnamefont{K.}~\bibnamefont{Ptaszy\ifmmode~\acute{n}\else
  \'{n}\fi{}ski}}, \bibinfo{journal}{Phys. Rev. B}
  \textbf{\bibinfo{volume}{95}}, \bibinfo{pages}{045306}
  (\bibinfo{year}{2017}).

\bibitem[{\citenamefont{Cao}(2006)}]{nonrenewal-cao}
\bibinfo{author}{\bibfnamefont{J.}~\bibnamefont{Cao}}, \bibinfo{journal}{The
  Journal of Physical Chemistry B} \textbf{\bibinfo{volume}{110}},
  \bibinfo{pages}{19040} (\bibinfo{year}{2006}).

\bibitem[{\citenamefont{Osad'ko and Fedyanin}(2011)}]{PhysRevA.83.063841}
\bibinfo{author}{\bibfnamefont{I.~S.} \bibnamefont{Osad'ko}} \bibnamefont{and}
  \bibinfo{author}{\bibfnamefont{V.~V.} \bibnamefont{Fedyanin}},
  \bibinfo{journal}{Phys. Rev. A} \textbf{\bibinfo{volume}{83}},
  \bibinfo{pages}{063841} (\bibinfo{year}{2011}).

\bibitem[{\citenamefont{Budini}(2010)}]{nonrenewal-budini}
\bibinfo{author}{\bibfnamefont{A.~A.} \bibnamefont{Budini}},
  \bibinfo{journal}{Journal of Physics B: Atomic, Molecular and Optical
  Physics} \textbf{\bibinfo{volume}{43}}, \bibinfo{pages}{115501}
  (\bibinfo{year}{2010}).

\bibitem[{\citenamefont{Witkoskie and Cao}(2006)}]{doi:10.1021/jp061471w}
\bibinfo{author}{\bibfnamefont{J.~B.} \bibnamefont{Witkoskie}}
  \bibnamefont{and} \bibinfo{author}{\bibfnamefont{J.}~\bibnamefont{Cao}},
  \bibinfo{journal}{The Journal of Physical Chemistry B}
  \textbf{\bibinfo{volume}{110}}, \bibinfo{pages}{19009}
  (\bibinfo{year}{2006}).

\bibitem[{\citenamefont{Saha et~al.}(2011)\citenamefont{Saha, Ghose, Adhikari,
  and Dua}}]{nonrenewal-enzyme}
\bibinfo{author}{\bibfnamefont{S.}~\bibnamefont{Saha}},
  \bibinfo{author}{\bibfnamefont{S.}~\bibnamefont{Ghose}},
  \bibinfo{author}{\bibfnamefont{R.}~\bibnamefont{Adhikari}}, \bibnamefont{and}
  \bibinfo{author}{\bibfnamefont{A.}~\bibnamefont{Dua}},
  \bibinfo{journal}{Phys. Rev. Lett.} \textbf{\bibinfo{volume}{107}},
  \bibinfo{pages}{218301} (\bibinfo{year}{2011}).

\bibitem[{\citenamefont{Cao and Silbey}(2008)}]{cao08}
\bibinfo{author}{\bibfnamefont{J.}~\bibnamefont{Cao}} \bibnamefont{and}
  \bibinfo{author}{\bibfnamefont{R.~J.} \bibnamefont{Silbey}},
  \bibinfo{journal}{The Journal of Physical Chemistry B}
  \textbf{\bibinfo{volume}{112}}, \bibinfo{pages}{12867}
  (\bibinfo{year}{2008}).

\bibitem[{\citenamefont{{Lang} and {Firsov}}(1963)}]{lang_firsov1963}
\bibinfo{author}{\bibfnamefont{I.~G.} \bibnamefont{{Lang}}} \bibnamefont{and}
  \bibinfo{author}{\bibfnamefont{Y.~A.} \bibnamefont{{Firsov}}},
  \bibinfo{journal}{Soviet Journal of Experimental and Theoretical Physics}
  \textbf{\bibinfo{volume}{16}}, \bibinfo{pages}{1301} (\bibinfo{year}{1963}).

\bibitem[{\citenamefont{Mitra et~al.}(2004)\citenamefont{Mitra, Aleiner, and
  Millis}}]{PhysRevB.69.245302}
\bibinfo{author}{\bibfnamefont{A.}~\bibnamefont{Mitra}},
  \bibinfo{author}{\bibfnamefont{I.}~\bibnamefont{Aleiner}}, \bibnamefont{and}
  \bibinfo{author}{\bibfnamefont{A.~J.} \bibnamefont{Millis}},
  \bibinfo{journal}{Phys. Rev. B} \textbf{\bibinfo{volume}{69}},
  \bibinfo{pages}{245302} (\bibinfo{year}{2004}).

\bibitem[{\citenamefont{Galperin et~al.}(2007)\citenamefont{Galperin, Ratner,
  and Nitzan}}]{galperin07}
\bibinfo{author}{\bibfnamefont{M.}~\bibnamefont{Galperin}},
  \bibinfo{author}{\bibfnamefont{M.~A.} \bibnamefont{Ratner}},
  \bibnamefont{and} \bibinfo{author}{\bibfnamefont{A.}~\bibnamefont{Nitzan}},
  \bibinfo{journal}{J. Phys.: Cond. Matt.} \textbf{\bibinfo{volume}{19}},
  \bibinfo{pages}{103201} (\bibinfo{year}{2007}).

\bibitem[{\citenamefont{Cuevas and Scheer}(2010)}]{moletronics}
\bibinfo{author}{\bibfnamefont{J.~C.} \bibnamefont{Cuevas}} \bibnamefont{and}
  \bibinfo{author}{\bibfnamefont{E.}~\bibnamefont{Scheer}},
  \emph{\bibinfo{title}{Molecular electronics: An introduction to theory and
  experiment}} (\bibinfo{publisher}{World Scientific}, \bibinfo{year}{2010}).

\bibitem[{\citenamefont{Srinivas and Davies}(2010)}]{Srinivas2010}
\bibinfo{author}{\bibfnamefont{M.}~\bibnamefont{Srinivas}} \bibnamefont{and}
  \bibinfo{author}{\bibfnamefont{E.}~\bibnamefont{Davies}},
  \bibinfo{journal}{Optica Acta: International Journal of Optics}
  \textbf{\bibinfo{volume}{28}}, \bibinfo{pages}{981} (\bibinfo{year}{2010}).

\bibitem[{\citenamefont{Zoller et~al.}(1987)\citenamefont{Zoller, Marte, and
  Walls}}]{Zoller1987}
\bibinfo{author}{\bibfnamefont{P.}~\bibnamefont{Zoller}},
  \bibinfo{author}{\bibfnamefont{M.}~\bibnamefont{Marte}}, \bibnamefont{and}
  \bibinfo{author}{\bibfnamefont{D.}~\bibnamefont{Walls}},
  \bibinfo{journal}{Physical Review A} \textbf{\bibinfo{volume}{35}},
  \bibinfo{pages}{198} (\bibinfo{year}{1987}).

\bibitem[{\citenamefont{Breuer and Petruccione}(2002)}]{Petruccione}
\bibinfo{author}{\bibfnamefont{H.~P.} \bibnamefont{Breuer}} \bibnamefont{and}
  \bibinfo{author}{\bibfnamefont{F.}~\bibnamefont{Petruccione}},
  \emph{\bibinfo{title}{The Theory of Open Quantum Systems}}
  (\bibinfo{publisher}{Oxford University Press, Oxford}, \bibinfo{year}{2002}).

\end{thebibliography}

\end{document}